\documentclass[aps,prl,amsmath,amssymb,twocolumn,showpacs,floatfix]{revtex4-1}
\usepackage{bm}
\usepackage{epsfig}
\usepackage[usenames]{color}
\usepackage{soul}
\usepackage{graphicx}
\usepackage[hypertex]{hyperref}
\usepackage{srctex}
\usepackage{pst-plot}
\usepackage{pstricks}
\renewcommand{\paragraph}[1]{\textit{#1.---} } 

\begin{document}

\title{Phase transitions in chiral magnets from Monte Carlo simulations }

\author{A.~M.~Belemuk}
\affiliation{Institute for High Pressure Physics, Russian Academy of Science, Troitsk 108840, Russia}
\author{S.~M.~Stishov}
\email{sergei@hppi.troitsk.ru}
\affiliation{Institute for High Pressure Physics, Russian Academy of Science, Troitsk 108840, Russia}
\begin{abstract}
Motivated by the unusual temperature dependence of the specific heat in MnSi, comprising a combination of a sharp first-order feature accompanied by a broad hump, we study the extended Heisenberg model with competing exchange $J$ and anisotropic Dzyaloshinskii-Moriya $D$ interactions in a broad range of ratio $D/J$. Utilizing classical Monte Carlo simulations we find an evolution of the temperature dependence of the specific heat and magnetic susceptibility with variation of $D/J$. Combined with an analysis of the Bragg intensity patterns, we clearly demonstrate that the observed puzzling hump in the specific heat of MnSi originates from smearing out of the virtual ferromagnetic second order phase transition by helical fluctuations, which manifest themselves in the transient multiple spiral state. These fluctuations finally condense into the helical ordered phase via a first order phase transition as is indicated by the specific heat peak. Thus the model demonstrates a crossover from a second-order to a first-order transition with increasing $D/J$. Upon further increasing $D/J$ another crossover from a first-order to a second-order transition takes place in the system. Moreover, the results of the calculations clearly indicate that these competing interactions are the primary factor responsible for the appearance of first order phase transitions in helical magnets with the Dzyaloshinskii-Moriya (DM) interaction.
\end{abstract}

\maketitle

\paragraph{Introduction}
The experimental and theoretical study of the model itinerant helimagnet MnSi is a dynamic area of active research  which still exhibits many unsolved problems \cite{a,1}. In particular, a tiny first order phase transition in MnSi and its accompanying phenomena present a real challenge that prevent a full understanding of the material. In this Letter we perform classical Monte Carlo simulations to get new insights on phase transitions in chiral spin systems.
As is known the crystal structure of MnSi belongs to the cubic space group P$2_{1}3$ , which does not contain a center of inversion.
This results in a non-zero value of the antisymmetric DM interaction term $D {\bf S}_i \times {\bf S}_j$. While the Heisenberg ferromagnetic interaction $J {\bf S}_i \cdot {\bf S}_j$ favors homogeneously aligned configurations of spins, the DM interaction leads to a long-wave modulation of the magnetic spin structure in a spiral type, appearing below a certain temperature $T_c$ ($T_c \approx 29$ K in MnSi) with a pitch wave vector ${\bf Q}$ directed along  the $[111]$  crystallographic direction \cite{1,2}. The magnetic moments are ferromagnetically aligned in planes perpendicular to $[111]$, and the planes are consecutively turned by some angle $\theta$ relative to each other, with $\tan \theta= D/J$.

\begin{figure}
\includegraphics[width=0.9\columnwidth]{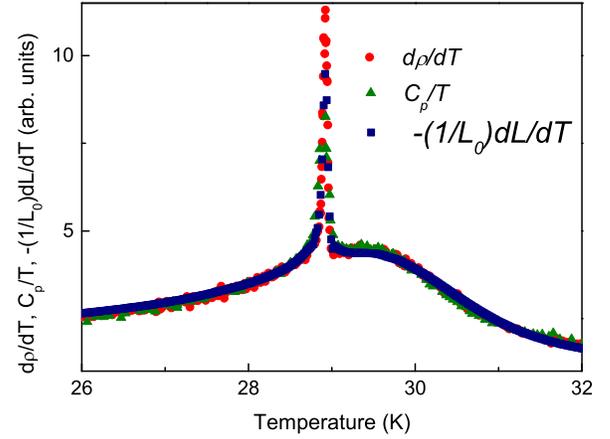}
\caption{(color online) Reduced specific heat divided by temperature $C_p/T$, linear coefficient of thermal expansion $1/L_0(dL/dT)$ and temperature derivative of resistivity $d\rho /dT$ (drawn after data of Ref.~\cite{3,4}).} \label{Fig1}
\end{figure}

There are strong indications that the magnetic transition in helimagnets is of first order \cite{2,3,4,5,6}. However, the nature of the phase transition in MnSi is still not well understood. This transition shows some remarkable features  seen in the specific heat, thermal expansion coefficient,  temperature coefficient of the resistivity and sound absorption  \cite{2,3,4,7},  As can be seen in Fig. \ref{Fig1} these quantities display sharp peaks at the phase transition temperature $T_c$ followed by  well-defined rounded humps on the high-temperature sides of the curves.

The origin of these enigmatic second maxima or humps is also poorly understood. Analyses of small angle neutron scattering data strongly advanced our knowledge, but still did not deliver completely satisfactory explanations \cite{6,8,9}.  What seems certain by now is that the domain of the secondary maxima is characterised by strong helical fluctuations evidenced by diffuse neutron scattering, with the intensity distributed on a sphere of radius ${\bf q}= {\bf Q}$ in momentum space \cite{6,8}.

Inspired by the successful simulation of a phase diagram of MnSi \cite{10}, we address the question on the nature of the phase transition in a chiral helimagnet by using a classical Monte Carlo technique. The chiral helimagnet is modelled by a lattice spin Hamiltonian consisting of Heisenberg exchange and varying DM interaction terms extended to include interactions between next-nearest neighbours.
This simple model leads to unusual magnetic properties based on the competing nature of the exchange and DM interactions. In particular, we show that the hump in the physical properties at the phase transition in MnSi originates in the perturbation of a virtual ferromagnetic second order phase transition by helical fluctuations that arise due to the DM interaction. In the other words the "hump" may be interpreted as a ferromagnetic second order phase transition smeared out by the helical fluctuations, which eventually condense into the helically ordered phase. This conclusion agrees completely with the analysis of the experimental thermodynamic data on the magnetic phase transition in MnSi, performed in \cite{11}.

\paragraph{Model  and simulation}
For a description of the itinerant helimagnet MnSi we adopt a local-moment model \cite{9,10,12,13}, whose Hamiltonian $H= H_J+ H_D$ consists of two terms, describing the exchange (J) and DM (D) interactions between nearest-neighbor spins, with DM vector directed along the corresponding bond direction,
\begin{widetext}
\begin{gather}
H_J= -J \sum \limits_{i} {\bf S}_i \cdot ({\bf S}_{i+ \hat x}+ {\bf S}_{i+ \hat y}+ {\bf S}_{i+ \hat z})- J' \sum \limits_{i} {\bf S}_i \cdot ({\bf S}_{i+ 2\hat x}+ {\bf S}_{i+ 2\hat y}+ {\bf S}_{i+ 2\hat z}), \label{HJ} \\
H_D= -D \sum \limits_{i}  \left({\bf S}_i \times {\bf S}_{i+ \hat x} \cdot \hat x+ {\bf S}_i \times {\bf S}_{i+ \hat y} \cdot \hat y+ {\bf S}_i \times {\bf S}_{i+ \hat z} \cdot \hat z \right)- D' \sum \limits_{i}  \left({\bf S}_i \times {\bf S}_{i+ 2\hat x} \cdot \hat x+ {\bf S}_i \times {\bf S}_{i+ 2\hat y} \cdot \hat y+ {\bf S}_i \times {\bf S}_{i+ 2\hat z} \cdot \hat z \right). \label{HD}
\end{gather}
\end{widetext}
We consider classical Heisenberg spins ${\bf S}_i= (S_i^x, S_i^y, S_i^z)$ of unit length, $|{\bf S}_i|= 1$ placed on a simple cubic lattice $L \times L \times L$ with periodic boundary conditions. The lattice spacing is taken to be unity.
The summation is over the sites of the cubic lattice spanned by the vectors $\hat x$, $\hat y$ and $\hat z$.
We also supplement exchange and DM interactions with next-nearest neighbor interaction terms $J'$ and $D'$ to compensate induced anisotropies originating from the discretization of the corresponding continuum spin model, as was first proposed by Buhrandt and Fritz \cite{10}.
Setting $J'= -J/16$ and $D'= -D/8$ allows to compensate these anisotropies to leading order \cite{10}.
In all calculations, except the one presented in Fig. \ref{Fig2} below, we held the parameter $J$ fixed at $J= 1$, serving as a unit of temperature.

To find thermodynamic properties we perform Monte Carlo (MC) simulations using a standard single-site Metropolis algorithm. The simulation starts at some temperature well above $T_c$ from some ordered state. During the equilibration stage of simulation it melts into paramagnetic disordered state. We then gradually decrease the temperature by small steps $\Delta T= 10^{-2}$. This enables us to obtain the dependence of specific heat as a function of temperature for varying DM strengths. Calculations were carried out with L= 30 and $5 \times 10^5$ MC steps per spin at each temperature to acquire the statistics.
The averaged lattice spin configuration $\langle {\bf S}_i \rangle$ at each $T$ is used to find
Fourier components, $\langle {\bf S}_q \rangle= 1/N\sum_i \langle {\bf S}_i \rangle e^{-i{\bf q}\cdot{\bf R}_i}$, where $N= L^3$, and, subsequently, the Bragg intensity profile $I({\bf q}) \propto |\langle {\bf S}_q \rangle|^2$. For ease of presentation, we also introduce the Bragg intensity profiles projected onto the ($q_x, q_y$) plane, $I^*(q_x, q_y)= \sum_{q_z} I(q_x,q_y,q_z)$.
Besides the Bragg intensity we have also studied the specific heat $C(T)$ and magnetic susceptibility
$\chi(T)= N(\langle {\bf M}^2 \rangle- \langle {\bf M} \rangle^2)/T$, where ${\bf M}= (1/N)\sum_i {\bf S}_i$.
We found it most convenient to calculate the specific heat from direct differentiation of the energy $E(T)= \langle H \rangle$.
We set $2\pi/L$ as a unit length in the reciprocal space and vectors ${\bf q}$ discussed below are scaled accordingly.

\begin{figure}
\includegraphics[width=1.1\columnwidth]{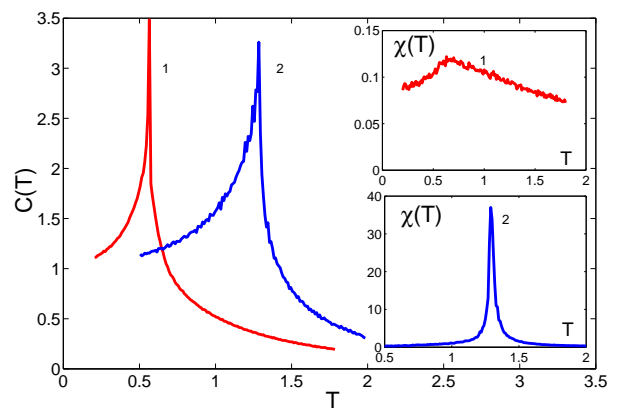}
\caption{(color online) Temperature dependence of the specific heat $C$ and magnetic susceptibility $\chi$ (insets)  for the DM Hamiltonian $H_D$ ($J= 0$, $D= 1$, curve 1) and the Heisenberg Hamiltonian $H_J$ ($J= 1$, $D= 0$, curve 2).} \label{Fig2}
\end{figure}
\begin{figure}

\includegraphics[width=1.05\columnwidth]{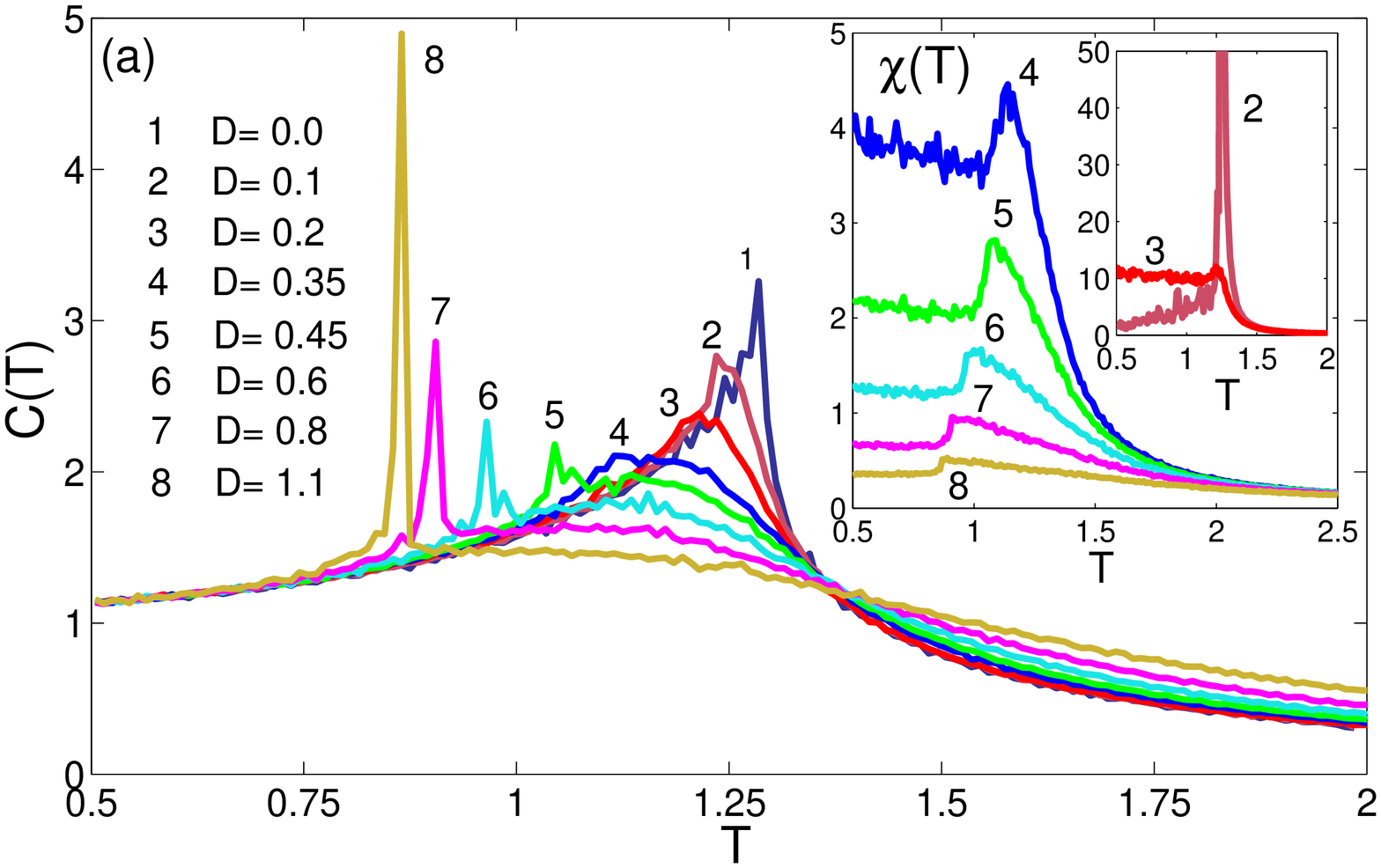}
\includegraphics[width=1.05\columnwidth]{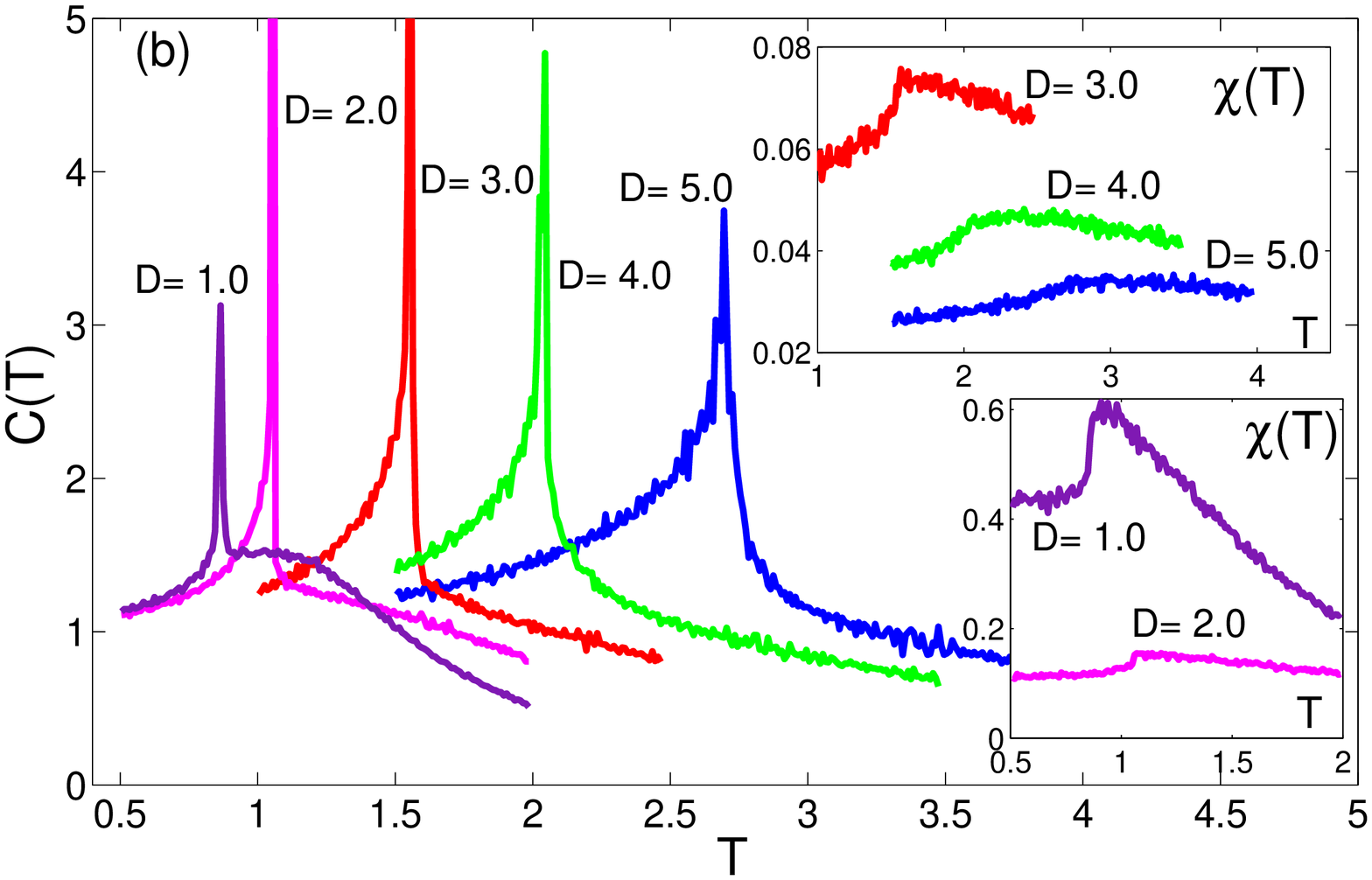}
\caption{(color online) (a),(b): Temperature dependence of the specific heat $C(T)$ for the Hamiltonian $H= H_J+ H_D$ for different values of the DM interaction. Inset: The corresponding dependence of the susceptibility $\chi(T)$. Note that a quasi invariant crossing point can be seen in panel (a) (see Refs. \cite{11,21}).} \label{Fig3}
\end{figure}
\begin{figure}
\includegraphics[width=\columnwidth]{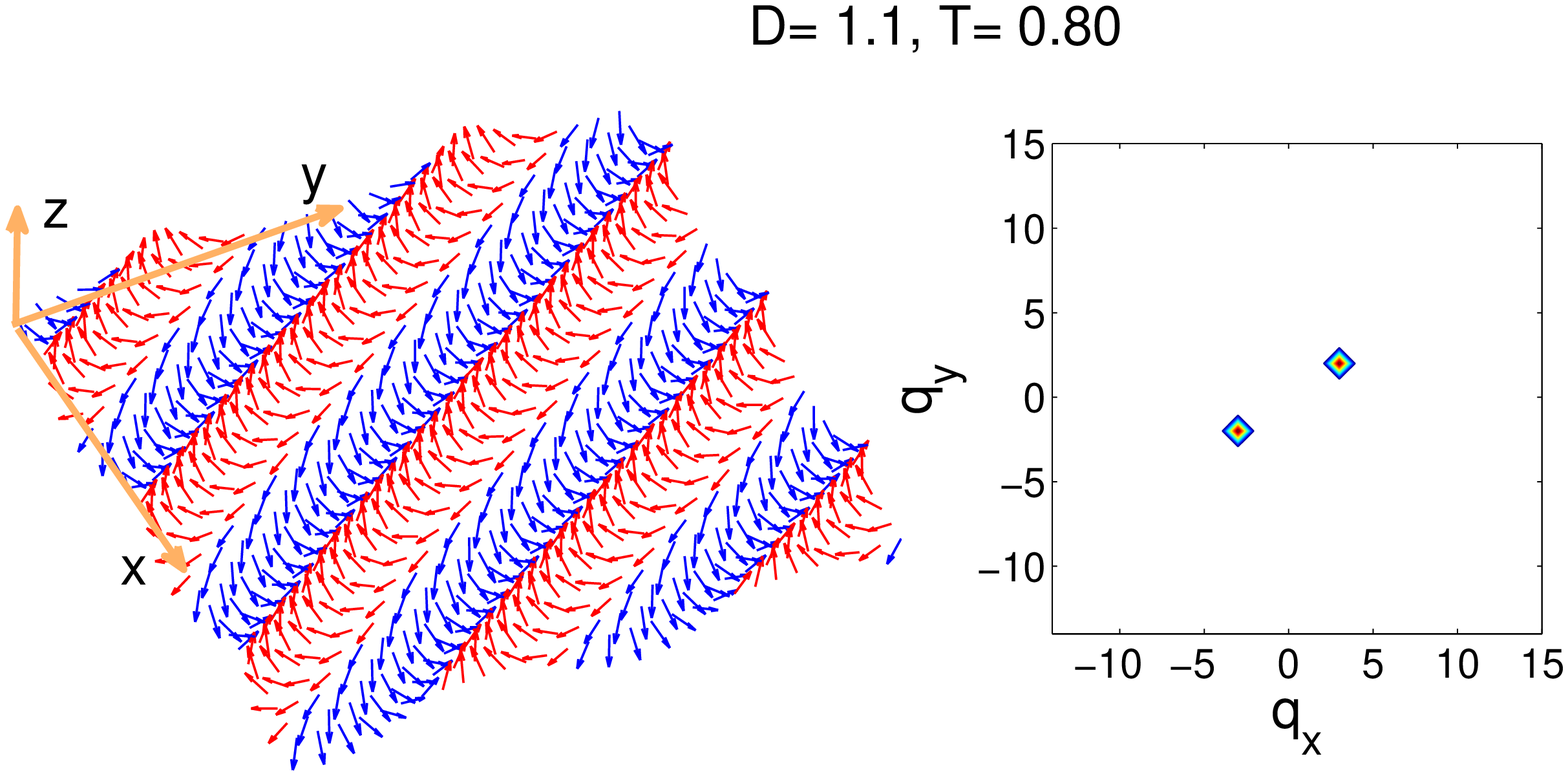}
\includegraphics[width=\columnwidth]{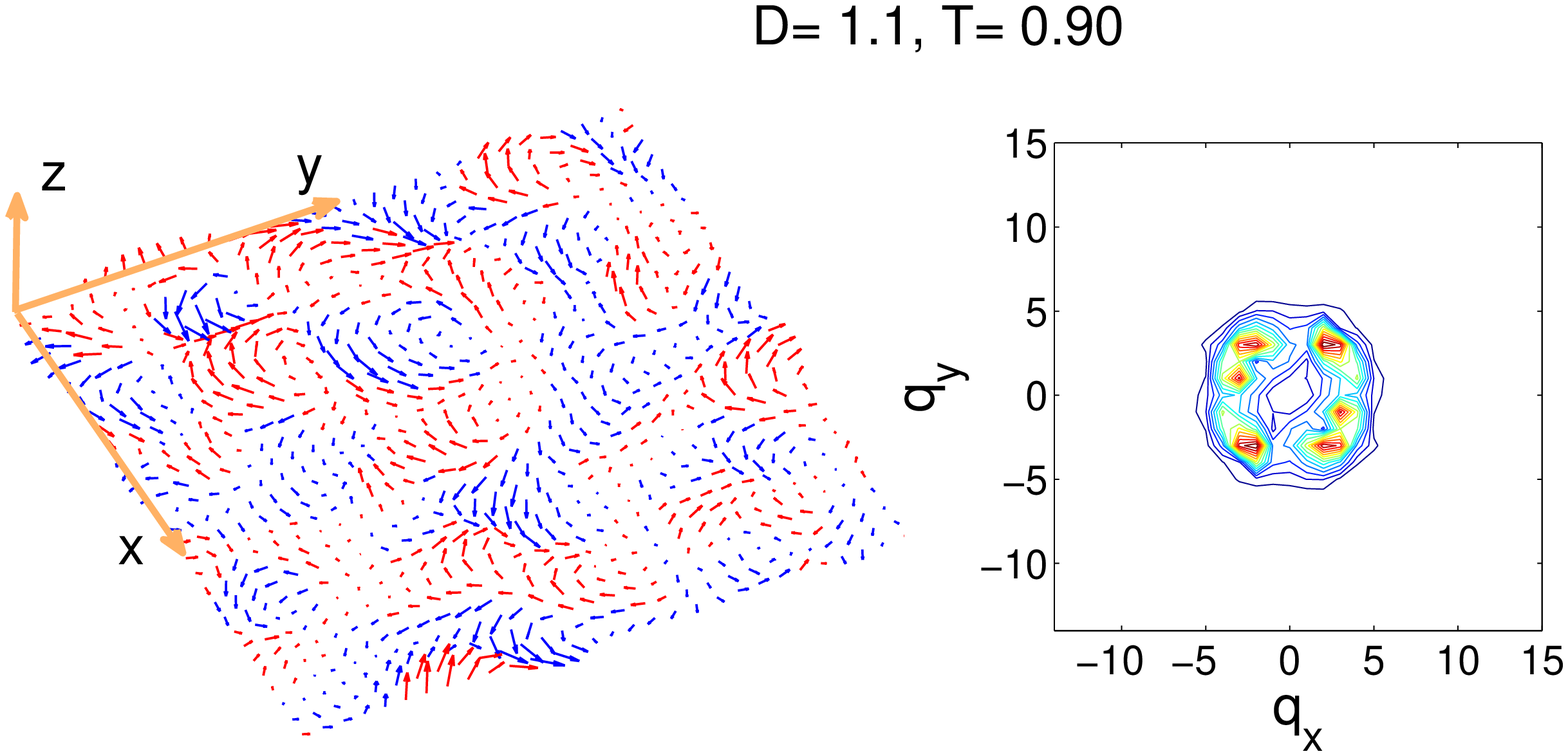}
\includegraphics[width=\columnwidth]{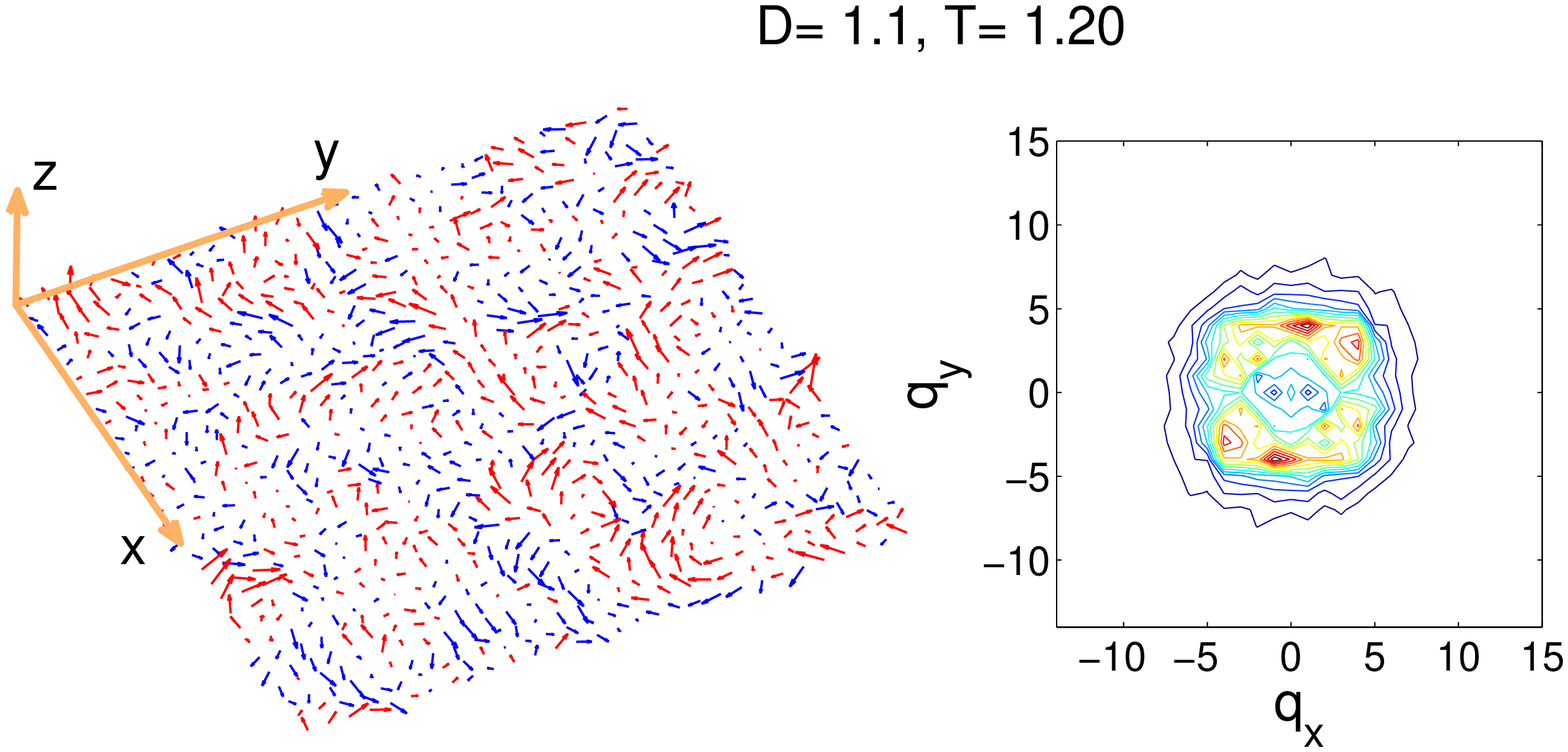}
\caption{(color online) Averaged spin configuration $\langle {\bf S}_i \rangle$ in the $xy$ plane (left) and the Bragg intensity profile $I^*$ projected onto the $(q_x, q_y)$ plane (right), both below and above the transition temperature $T_c \simeq 0.86$ for DM interaction $D= 1.1$. Spins with positive (negative) values of $S_z$ are presented in red (blue). The size of the arrows is proportional to $|\langle {\bf S}_i \rangle|$. Distances in reciprocal space are scaled by $2\pi/L$.} \label{Fig4}
\end{figure}

\begin{figure}
\includegraphics[width=\columnwidth]{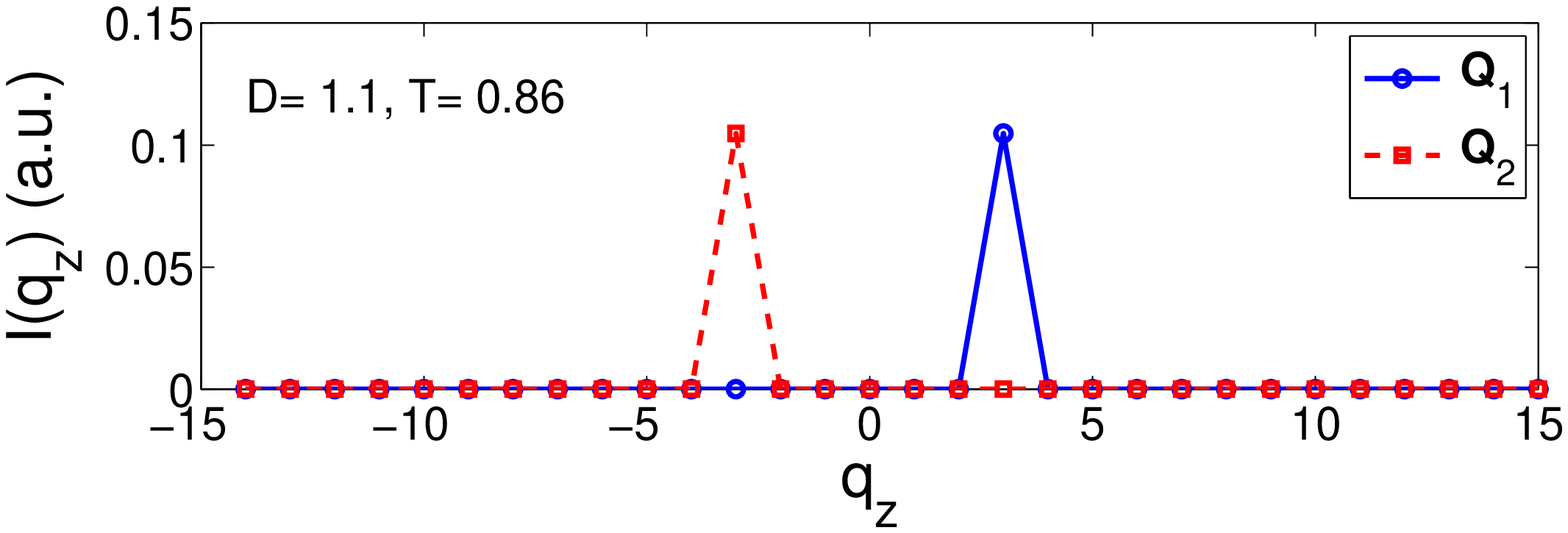}
\includegraphics[width=\columnwidth]{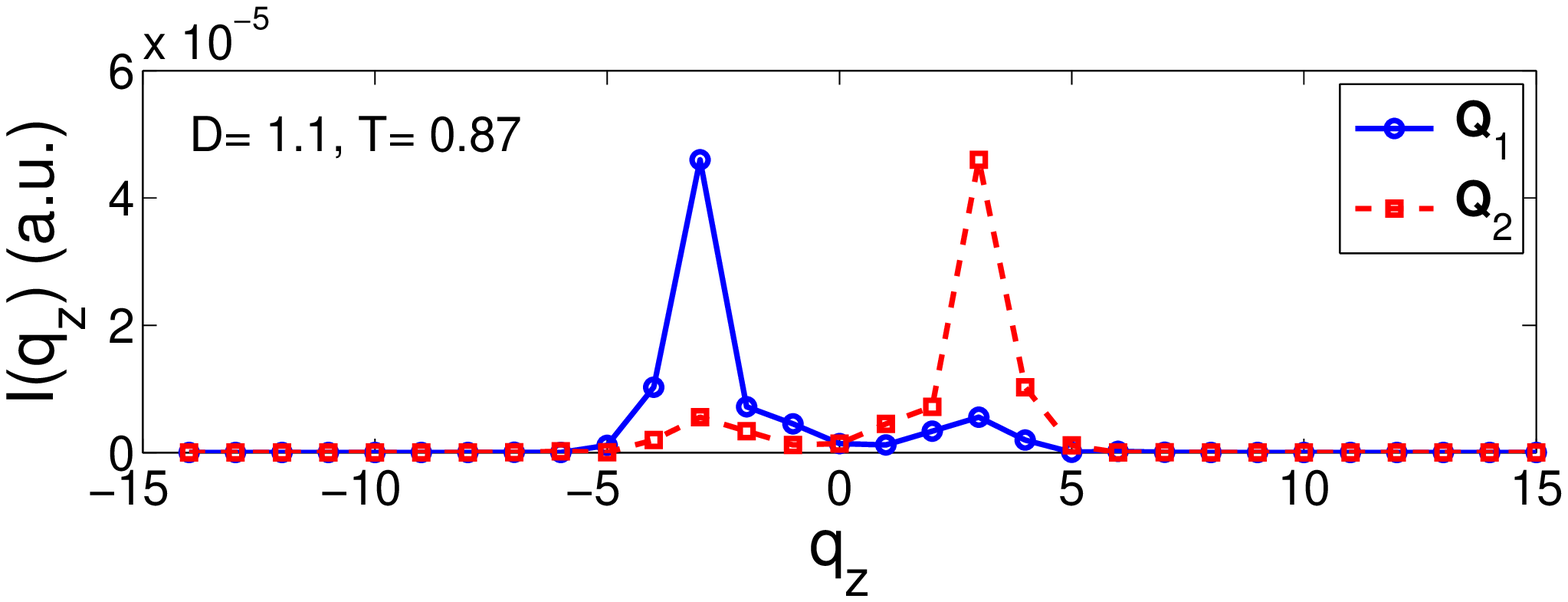}
\includegraphics[width=\columnwidth]{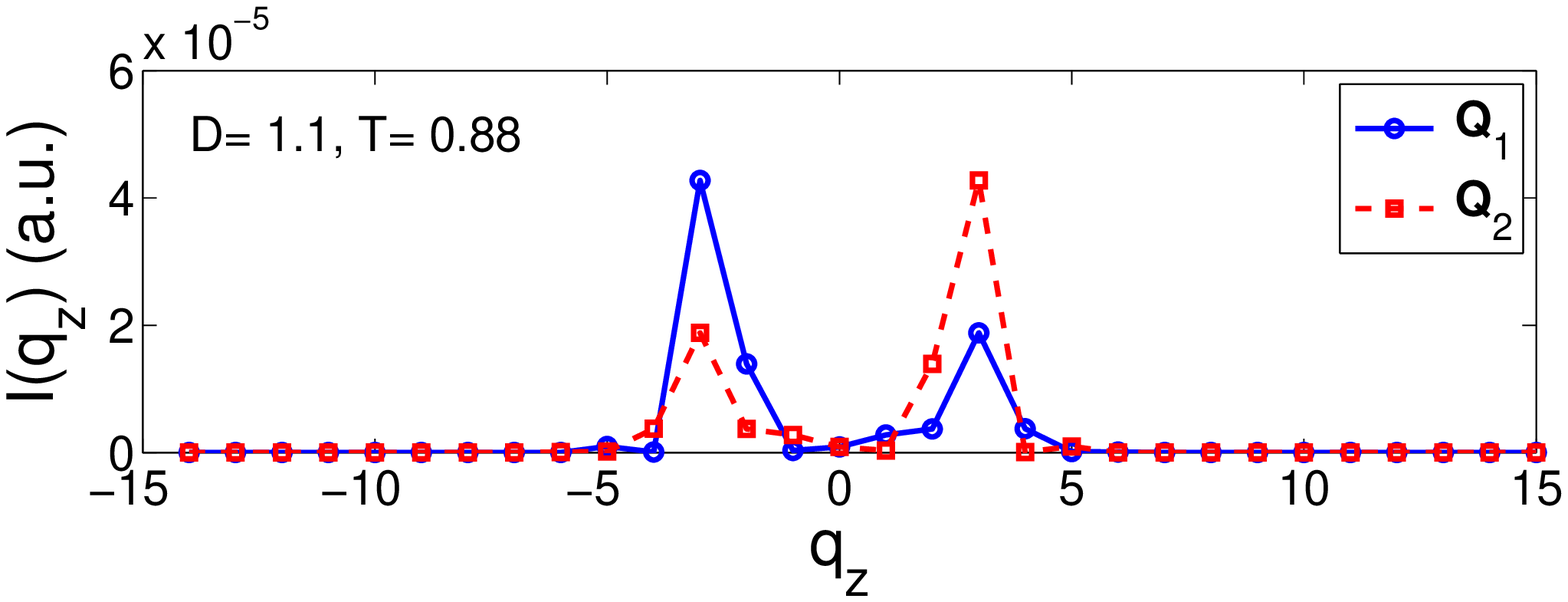}
\caption{(color online) Intensity profile of $I(q_x, q_y, q_z)$ near the transition temperature ($T_c= 0.86$, D= 1.1) as a function of $q_z$ taken at $(q_x, q_y)= {\bf Q}_1$ and $(q_x, q_y)= {\bf Q}_2$. The two-dimensional vectors ${\bf Q}_1= (3, 2)$ and ${\bf Q}_2= (-3, -2)$ correspond to the two peaks in $I^*(q_x, q_y)$ at $T < T_c$ shown in the upper panel of Fig. \ref{Fig4}. Values of $q_z$ are scaled by $2\pi/L$.} \label{Fig5}
\end{figure}

\paragraph{Results and discussion}
The Hamiltonian $H$ contains two main terms, the Heisenberg interaction term $H_J$, and the antisymmetric DM intraction $H_D$. Separately, each of them results in a phase transition of second order at some $T_c$ from an ordered phase to a disordered paramagnetic phase. The behavior of the specific heat and magnetic susceptibility for these two transitions is illustrated in Fig. \ref{Fig2} (in both cases the compensation terms $J'$ and $D'$ are included).
The corresponding susceptibility for the Hamiltonian $H_D$ (curve 1 in the inset of Fig. 2) shows an antiferromagnetic signature, whereas Hamiltonian $H_J$ leads to a singularity in the susceptibility at $T= T_c$  (curve 2 in the inset of Fig. 2) as it expected for  ferromagnets.

The specific heat shown in Fig. \ref{Fig2} (curve 1 and 2) displays the singular behavior at $T_c$ typical of second order transitions.
In  case of the Heisenberg interaction the ordered ferromagnetic (FM) state is characterized by a single peak in Bragg intensity at zero wave vector. With increasing temperature this peak gradually vanishes signalizing the disappearance of the ordered FM state, at a transition temperature $T_c/ J \simeq 1.44$ \cite{14}. In our simulation the transition occurs at slightly lower temperature, $T_c \simeq 1.28$ due to finite size effects.

In the case of DM interactions the lower-temperature phase is characterized by a spiral magnetic order
${\bf S}({\bf R}_i)= S_Q \cos ({\bf Q}\cdot {\bf R}_i) {\bf e}_1+ S_Q \sin ({\bf Q}  \cdot {\bf R}_i) {\bf e}_2$, with unit vectors ${\bf e}_{1}$ and ${\bf e}_{2}$ being perpendicular to the spiral vector ${\bf Q}$.
Such a single spiral results in two Bragg peaks at ${\bf Q}$ and $-{\bf Q}$. In our simulation the corresponding spiral wave vector turns out to be ${\bf Q}= (2\pi/L)(9,9,9)$, which corresponds to relative rotations of spins in neighboring planes perpendicular to $[111]$ by an angle approximately equal to $\pi/2$ (generally, the magnitude of the spiral vector is affected by the periodic boundary conditions and the competing term $D'$).
The eventual transition to the paramagnetic state, accompanied by simultaneous diminishing of the Bragg peaks' intensity, occurs at $T_c/D \simeq 0.56$.

For the combined Hamiltonian $H= H_J+ H_D$ the competing nature of the terms $H_J$ and $H_D$, which have different ground states, results in strong fluctuations that drive the system from a second-order to a first order transition at some values of $D/J$. The hump or shoulder in the specific heat at the high temperature side of the phase transition appears to result from a degradation of the ferromagnetic second order phase transition caused by the intense helical fluctuations.
To reach this conclusion we performed a number of runs, varying the parameter $D$, while $J= 1$ being fixed.
The evolution of the specific heat and susceptibility for small and large values of $D$ are presented in panels (a) and (b) of Fig. \ref{Fig3}. At $D < 0.3$, the specific heat exhibits a clear maximum, resembling the singular behavior of $C(T)$ at $T_c$ for $D= 0$. With increasing $D$ this maximum evolves into a round hump, which eventually disappears for $D > 1$. Concurrently, a first order sharp peak of $C(T)$ develops at the low temperature side of the hump. Upon further  increasing $D$ beyond $D = 1$ the phase transition temperature $T_c$ also increases while simultaneous changing the shape of the $C(T)$-curve.  The transition becomes second order again. Examining the corresponding dependencies of the susceptibility $\chi(T)$ reveals that with increasing  $D$ the maximum of $\chi(T)$ evolves into a step-like feature with considerably diminished magnitude of $\chi$. At $D \geq 3$ the susceptibility is  almost featureless with a nearly vanishing magnitude.

To examine the corresponding real-space spin structures and their counterparts in reciprocal space we show in Fig. \ref{Fig4}  changes of the averaged spin configuration $\langle {\bf S}_i \rangle$ and Bragg intensity $I^*(q_x, q_y)$ as function of temperature. For illustration we chose the value of $D= 1.1$ where the transition temperature $T_c \simeq 0.86$
(for $T > T_c$ the averaged spin at a given site $\langle {\bf S}_i \rangle$ becomes very small, and we have to re-scale spin lengths by some factor for presentation purposes).

For temperatures below the transition temperature one can clearly see a spiral state (upper panel), with a two-peak structure of the Bragg intensity profile at (two-dimensional) vectors ${\bf Q}_1= (2\pi/L) (3,2)$ and ${\bf Q}_2= (2\pi/L) (-3,-2)$. The corresponding spiral vector is ${\bf Q}= (2\pi/L) (3,2,3)$.
When increasing $T$ just above $T_c$ the spin pattern changes drastically (middle panel): one can see an emergence of the four-peak structure. This four-peak structure gradually transforms into a ring-shape form (bottom panel). A similar intensity patterns is also observed when projecting onto the $(q_x, q_z)$ plane. A close examination of the spin configuration in real space shows that there are sites with vanishing values of averaged spin $\langle {\bf S}_i \rangle$. Configurations with $\langle {\bf S}_i \rangle= 0$ arise in  multi-spiral states predicted for 2D models with DM interaction \cite{13, 15, 16, 17}, or for competing frustrated Heisenberg models on triangular lattices \cite{18}.

To take a close look at this multi-spiral spiral state we present $I({\bf q})$ as function of $q_z$ in Fig. \ref{Fig5}.
The data are shown for $D= 1.1$ at three temperatures around the transition temperature $T_c= 0.86$. The one-spiral state is easily identified at $T= T_c$ (upper panel), where peaks at $q_z= 3$ and $q_z= -3$ correspond to ${\bf Q}$  and $-{\bf Q}$.
Slightly above $T_c$, (middle and bottom panels) the system starts to develop two main peaks at ${\bf Q}_1$ and ${\bf Q}_2$,
which signals the emergence of a quasi two-spiral state. With increasing $T$ other peaks emerge and this quasi two-spiral state gradually transforms into the ring-shape structure seen in Fig. \ref{Fig4} (bottom panel).

With increasing  $D$ the multi-spiral state transforms into a ring-shape state of very small intensity $I({\bf q})$, which  subsequently fades away.
This nicely correlates with a disappearance of the hump in $C(T)$ as $D$ increases.
So it is important to emphasize that the spin configuration in the hump temperature region corresponds to the above-mentioned multi-spiral structure, which melts away with an increase in temperature or $D$.

\begin{figure}
\includegraphics[width=\columnwidth]{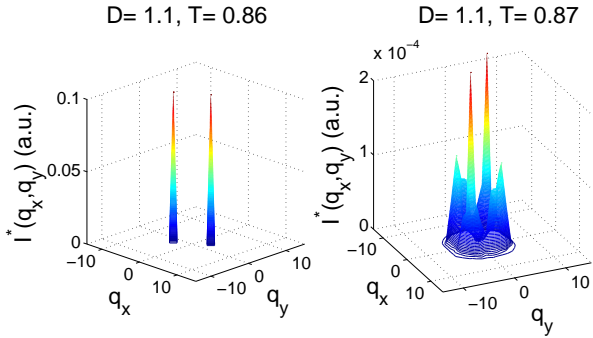}
\caption{(color online) Projected Bragg intensity $I^*(q_x, q_y)$ at $T= T_c$ (left) and slightly above $T_c$ (right) for DM interaction D= 1.1. Values of $q_x$, $q_y$ are scaled by $2\pi/L$.} \label{Fig6}
\end{figure}

\begin{figure}
\includegraphics[width=\columnwidth]{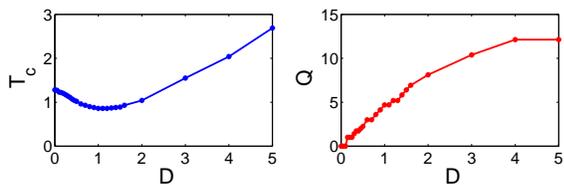}
\caption{(color online) Transition temperature $T_c$ (left) and magnitude of the spiral vector $Q$ of the low-temperature phase (right) as function of DM interaction strength. Length of $Q$ is scaled by $2\pi/L$. } \label{Fig7}
\end{figure}

To better understand the phase transition studied here we show $I(\bf{q})$ in Fig. \ref{Fig6}, which illustrates the drastic change  at the transition temperature for the case $D= 1.1$. The left panel shows $I({\bf q})$ at $T_c$, and the right panel just above it. The intensity profile at $T_c$ displays the spiral state with peaks at the spiral vector ${\bf Q}= 2\pi/L (3,2,3)$ and at $-{\bf Q}$.
It is seen that the intensity drops by nearly three orders of magnitude just above $T_c$. It results from the corresponding drop in the magnitude of $\langle {\bf S}_i \rangle$ and is a clear signature of a first order transition. A similar behavior of $I({\bf q})$ is observed over a large $D$ range, $0.3 < D < 2$. This range for $D$ also covers the region where  $T_c(D)$ has its minimum, as shown in Fig. \ref{Fig7}. Upon further increase of $D$, the discontinuous jump in $I({\bf q})$ at $T_c$ gradually vanishes, signaling that the transition becomes second order again.

The full dependence $T_c(D)$ is shown in Fig. \ref{Fig7}. The data points around the minimum of $T_c(D)$ correspond to the first order transition in the system. Away from this region, when $T_c$ gradually grows, the transition becomes second order. Looking again at Fig. \ref{Fig6} it is tempting to consider the ratio $D/J$ as a parameter that could control the evolution of the phase transition in the helical magnets at high pressures. This supposition can probably be valid for systems with local magnetic moments (for instance, in the case of Cu$_2$OSeO$_3$ \cite{19}), but not directly applicable to an itinerant system like MnSi. However, a variation of the ratio $D/J$ could be an important factor influencing the phase diagrams of various materials.
Another factor that can govern the phase diagram is a magnitude of compensating terms $J'$ and $D'$ in the Hamiltonian $H$. In our case these terms slightly shift a position of the first-order peak in the dependence of $C(T)$ to a lower temperature, with the spin structure of the hump being not affected.

In Fig. \ref{Fig7} we also show the $D$-dependence of the magnitude $Q$ of the spiral vector in the ordered low-temperature phase. With increasing $D$ the spiral wave vector continuously increases, while simultaneously rotating from direction $[100]$ via $[110]$ to the body diagonal $[111]$ to better adjust the spiral to the structure of the lattice.

Finally,  classical Monte Carlo simulations of a spin system with the competing interactions $J$ and $D$ reproduce very well the experimental situation observed in helical magnets MnSi and Cu$_2$OSeO$_3$. The hump or shoulder in the specific heat on the high temperature side of the first order peak arises as a consequence of a perturbation of the virtual second order ferromagnetic phase transition by helical fluctuations. The hump domain therefore has a very complicated spin structure (for illustration see Fig.~ \ref{Fig4}) stipulated by the competing ferromagnetic and helical fluctuations. As a result, at lower temperature the system cannot proceed to an ordered state in a continuous way and takes on helical order through a first order transition. We therefore propose that the competing interactions are the primary factor responsible for an occurrence of first order phase transitions in helical magnets with the Dzyaloshinskii-Moriya interaction. Indeed, with two competing interactions and two interacting order parameters in a system one may expect a renormalisation and a sign change of the forth order term in the Landau expansion, which would lead to a first order phase transition. The arising spin fluctuations (see \cite{6,20}) are a direct consequence of this situation.

\paragraph{Conclusion}
Summarizing our results we found strong evidence for a continuous transformation of a second-order transition into a first-order transition with varying $(D/J)$. At low and high $D$ the system is perturbed weakly from the respective ground state of $H_J$ or $H_D$. At intermediate values of $(D/J) \sim 1$ the system exhibits strong competing interactions resulting in a continuous change of the nature of the transition from second to first-order, with subsequent reentrant behavior when at large $D \gtrsim 2$ the system again exhibits second-order behavior. We show that the hump in the physical properties at the phase transition in MnSi and Cu$_2$OSeO$_3$ originates from a perturbation of a virtual ferromagnetic second order phase transition by helical fluctuations arising due to the DM interaction. In the other words the "hump" may be viewed as a smeared out ferromagnetic second order phase transition by the helical fluctuations which eventually condense into the helical ordered phase. This conclusion completely agrees with the assumption formulated in \cite{11}.

\paragraph{Acknowledgements}
AMB (calculations, data analysis, writing the paper) and SMS (data analysis,writing the paper) greatly appreciate financial support   of  the Russian Foundation for Basic Research (grant 15-02-02040, 17-52-53014), Program of the Physics Department of RAS on Strongly Correlated Electron Systems and Program of the Presidium of RAS on Strongly Compressed Matter.


\begin{thebibliography}{99}



\bibitem{a}	M.Brando, D.Belitz, F.M. Grosche, T.R. Kirkpatrick, Rev.Mod.Phys. {\bf 88}, 025006 (2016).
\bibitem{1}	S.M. Stishov, A.E. Petrova, Physics-Uspekhi, \textbf{11}, 1117 (2011)
\bibitem{2}	Y. Ishikawa, K. Tajima, D. Bloch, M. Roth, Solid State Commun. {\bf 19}, 525 (1976).
z\bibitem{3}	S. M. Stishov, A. E. Petrova, S. Khasanov, G. K. Panova, A. A. Shikov, J. C. Lashley, D. Wu, T. A. Lograsso, Phys. Rev.,\textbf{B 76}, 052405 (2007).
\bibitem{4}	S. M. Stishov, A. E. Petrova, S. Khasanov, G. K. Panova, A. A. Shikov, J. C. Lashley, D. Wu, T. A. Lograsso, J. Phys.: Condens. Matter,\textbf {20}, 235222  (2008).
\bibitem{5}	C.Pappas, E. Leli\'{e}vre-Berna, P. Falus, P. M. Bentley, E. Moskvin, S. Grigoriev, P. Fouquet, B. Farago, Phys. Rev. Lett.,\textbf {102}, 197202 (2009).
\bibitem{6} M. Janoschek, M. Garst, A. Bauer, P. Krautscheid, R. Georgii, P. B\"{o}ni, C. Pfleiderer, Phys. Rev. B {\bf 87}, 134407 (2013).
\bibitem{7} A. E. Petrova, S. M. Stishov, J. Phys.: Condens. Matter {\bf 21}, 196001 (2009).
\bibitem{8}	S.V. Grigoriev, S.V. Maleyev, E.V. Moskvin, V.A. Dyadkin, P. Fouquet and H. Eckerlebe, Phys. Rev.,\textbf{B 81}, 144413 (2010).
\bibitem{9}	A. Hamann, D. Lamago, Th. Wolf, H. v. L\"{o}hneysen and D. Reznik, Phys. Rev. Lett.,\textbf{107}, 037207 (2011).

\bibitem{10} S. Buhrandt and L. Fritz, Phys. Rev. B {\bf 88}, 195137 (2013).
\bibitem{11} S. M. Stishov, A. E. Petrova, Phys. Rev. B {\bf 94}, 140406(R) (2016).
\bibitem{12} S. D. Yi, S.Onoda, N.Nagaosa, J.Hoon Han, Phys. Rev. B {\bf 80}, 054416 (2009).
\bibitem{13} A. O. Leonov and M. Mostovoy, Nat. Commun, 6:8275 (2015).

\bibitem{14} R. G. Brown and M. Ciftan, Phys. Rev. Lett. {\bf 76}, 1352 (1996).

\bibitem{15} M. Ezawa, Phys. Rev. B {\bf 83}, 100408 (2011).
\bibitem{16} J.-H. Park and J. H. Han, Phys. Rev. B {\bf 83}, 184406 (2011).
\bibitem{17} B. Binz, A. Vishwanath, and V. Aji, Phys. Rev. Lett. {\bf 96}, 207202 (2006).
\bibitem{18} T. Okubo {\it et al.}, Phys. Rev. Lett. {\bf 108}, 017206 (2012).
\bibitem{19} V. A. Sidorov, A. E. Petrova,  P. S. Berdonosov, V. A. Dolgikh,  and S. M. Stishov, Phys. Rev. B {\bf 89}, 100403(R) (2014).

\bibitem{20}	P. Bak, M. H\o gh Jensen, J. Phys C: Solid St. Phys.,\textbf{13}, L 881 (1980).

\bibitem{21} D. Vollhardt, Phys. Rev. Lett. {\bf 78}, 1307 (1997).


\end{thebibliography}
\end{document}